\documentclass[12pt]{article}
\usepackage[dvips]{graphicx}
\usepackage{epsfig}
\usepackage{afterpage}
\usepackage{amsmath}
\usepackage{amssymb}
\begin{document}
 \begin{center}
 {\large\textbf
      {Inflationary Brans-Dicke Quantum Universe II:
       Particular evolutions and stability analysis}}
 \end{center}
 \vspace{1mm} %
\begin{center}
 B. Geyer \\
 \vspace{1mm} %
{\footnotesize\textit{ Center of Theoretical Studies, Leipzig
University, Augustusplatz 10, 04109 Leipzig, GERMANY; \\ e-mail:
geyer@itp.uni-leipzig.de}}
\\ \vspace{2mm} and \\
V. F. Kovalev\\
\vspace{1mm}%
{\footnotesize\textit{ Institute for Mathematical Modelling,
Miusskaya Sq., 4a, 125047, Moscow, RUSSIA; \\ e-mail:
kovalev@imamod.ru}}
\\
\end{center}
\vspace{1mm}

\begin{abstract}
We made an analysis of the equations of motion which are obtained
from the one--loop effective action for Brans--Dicke gravity with
$N$ dilaton--coupled massless fermions in a time--dependent
conformally flat background \cite{GOZ}. Various particular
solutions, including the well-known stationary one, of the
corresponding set of first-order differential equations are given.
Some of these solutions describe an expanding time-dependent
Universe with increasing, constant or also decreasing dilaton.
This is illustrated by a numerical analysis. For the nonstationary
solutions a stability analysis is given.
\end{abstract}

\numberwithin{equation}{section}


\section{Introduction}

In recent times Brans-Dicke(BD) gravity with matter in the
Einstein frame attracted much attention. Especially, such theories
have been realized as Einstein gravity with dilaton coupled to
matter (for a review of the renewed interest in scalar--tensor
gravity see Ref.~\cite{FGT}). Obviously, Brans-Dicke theory
\cite{will} represents one of the simplest examples of
scalar--tensor (or dilatonic) gravities where the background is
described by the metric and the dilaton. The reason for the
consideration of such models is the following:
\\
First of all, the dilaton is an essential element of string
theories, and the low--energy string effective action  may be
considered as some kind of BD theory with higher order terms (for
a recent review, see \cite{polch}).
%\\
Second, dimensional reduction of Kaluza-Klein theories naturally
may lead to BD gravity.
%\\
Third, dilatonic gravity is expected to have such important
cosmological applications as, e.g., in the case of (hyper)extended
inflation
 \cite{la}. In addition, there was some activity on
the study of BD cosmologies with varying speed of light \cite{barrow}.

Recently, the anomaly-induced action for dilaton coupled scalars,
vectors and spinors in four dimensional curved spacetime has been
calculated in Ref.~\cite{DEA}. Thereafter, in \cite{GOZ}, the
effective action formalism has been applied for a study of quantum
cosmology in BD gravity with dilaton coupled spinors. There, the
one-loop anomaly--induced effective action (EA) due to $N$
massless fermions on a time-dependent conformally flat background
coupled to the dilaton has been computed and the (fourth-order)
quantum--corrected equations of motion have been derived. However,
because of the complicated structure of that coupled set of
(ordinary) differential equations only one special solution,
representing an inflationary Universe with slowly expanding BD
dilaton, could be presented.

Here, we study the equations of motion, for conformal as well as
physical time, more extensively. We were able to show additional
particular, physically relevant solutions. Making use of the
symmetry group analysis we showed that the former solution is a
special case of the "stationary solution" of the set of five
first-order differential equations being equivalent to the
equations of motion. By a numerical analysis the asymptotic
behaviour of various solutions will be shown; some of them
approximate the stationary solution. For the non-stationary
solutions a perturbative analysis has been done and some new
solutions corresponding to terminating series are presented.
Finally, a stability analysis of the equations of motion is given.

\section{Equations of motion: conformal time}

The action (in the Einstein frame) of BD theory with dilaton $\phi$
 coupled to $N$ massless Dirac spinors is \cite{GOZ}
 \begin{equation}
  S=\int \rm d^4x \,\sqrt{-g} \left[\frac{R}{16\pi G} -\frac{1}{2}
 (\nabla_\mu \phi)(\nabla^\mu \phi)+ \exp{(A\phi)}\sum_{i=1}^N
 \bar\psi_i\gamma^\mu\nabla_\mu\psi^i  \right] \,,
 \label{bde1}
 \end{equation}
with $A=-8\sqrt{\pi G/(2\omega+3)}, \  \omega > -3/2$; $G$ is
Newtons constant and $\omega$ is the BD coupling parameter of the
dilaton. In the following we restrict ourselves to FRW type
cosmologies,
 \begin{equation}
 \textrm{d} s^2=-\textrm{d} t^2+a^2(t) \textrm{d} l^2\,, \label{st4}
 \end{equation}
where $dl^2$ is the line metric element of a 3-dimensional space
$\Sigma$ with constant curvature.
It is convenient to introduce conformal time $\eta$ by means of
 \begin{equation}
 \label{conftime}
 \textrm{d} t= a(\eta) \textrm{d} \eta\,,
 \end{equation}
to get a space--time which is conformally related to an ultrastatic
space--time with spatial section $\overline{M}$ of constant curvature.
For the flat case ($\Sigma=R^3$, i.e., $k=0$) the metric gets
 \begin{eqnarray}
 g_{\mu \nu} = a^2(\eta) \bar{g}_{\mu \nu}\, \qquad
 \textrm{with}\qquad \bar{g}_{\mu \nu}\equiv \eta_{\mu\nu}, \qquad
 a(\eta)=\textrm{e}^{\sigma(\eta)}\,. \nonumber
 \end{eqnarray}

In that case the complete anomaly--induced EA for the dilaton coupled
spinor field becomes \cite{GOZ}
 \begin{equation}
 \label{1ef}
 W=V_3\int \textrm{ d} \eta \left\{ b_2 \tilde\sigma
{\tilde\sigma}''''  - (b_1 + b_2){\left( \tilde\sigma'' -
 (\tilde\sigma'){}^{2} \right)}^2 \right\} \ ,
 \end{equation}
where $V_3$ is the (infinite) volume of flat 3--space,
$\tilde\sigma=\sigma+ A\phi /3$ ($'\equiv{\textrm{d} /
\textrm{d}\eta}$) and, for Dirac spinors, $b_1={N / 10(4\pi)^2}$,
$b_2=-{11 N / 180 (4\pi)^2}$. The total one--loop EA is obtained
by adding to $W$ the classical action (for $k=0$):
 \begin{equation}
 S= \frac{1}{2}\, V_3\int \textrm{ d}\eta
 \left[ \frac{3}{4 \pi G}(\sigma''+\sigma'{}^2)
 + \phi'{}^2\right] \textrm{e}^{2\sigma}\,.
 \label{ca}
 \end{equation}

Then, the corresponding equations of motion are (see Eqs. (15) and
(16) of \cite{GOZ})
 \begin{equation}
  \label{GOZ1}
 {\widetilde C} \textrm{e}^{A\phi/3} + ({3}/{4 \pi G})a'' +
 a {\phi}'{}^2 = 0\, ,
 \end{equation}
 \begin{equation}
  \label{GOZ2}
  ({A}/{3}){\widetilde C} a \textrm{e}^{A\phi/3}-(a^2\phi')'  =  0\,,
 \end{equation}
where
 \begin{equation}
 \label{GOZ3}
 {\widetilde C} = - \frac{2b_1}{\tilde a}
 \Bigg[\left(\frac{{\tilde a}'}{{\tilde a}}\right)''
 -2 \left(1+\frac{b_2}{b_1}\right)
 \left(\frac{{\tilde a}'}{{\tilde a}}\right)^3\Bigg]'
 \end{equation}
 with the notations
 \[
 {\tilde a}(\eta) \equiv {\textrm{e}}^{\tilde\sigma(\eta)} =
 a(\eta)\, {\textrm{e}}^{v(\eta)}, \qquad v(\eta) = A\phi(\eta)/3.
 \]
A complete integration of these equations appears to be hopeless.
Therefore, let us ask for particular solutions.

First of all, Eq.~(\ref{GOZ2}) because of (\ref{GOZ3}) may be integrated
immediately. Then the system (\ref{GOZ1}) -- (\ref{GOZ3}), written
in terms of the variables $\tilde a$ and $v$, reads:
 \begin{equation}
 \label{basic1}
 (1-\alpha) [(v'' - v'{}^2){\tilde a} +
 2{\tilde a}'v'] + \alpha {\tilde a}''  =  0\,,
 \end{equation}
  \begin{equation}
  \label{basic2}
 {\tilde a}^2{\tilde a}''' - 3{\tilde a}{\tilde a}'{\tilde a}''
 - 2\nu {\tilde a}'{}^3 +\beta {\tilde a}^5 v'{\textrm e}^{-2v}  = B{\tilde
 a}^3\,,
 \end{equation}
where $B$ is an arbitrary integration constant, and the following
notations have been used:
 \[
  \alpha \equiv \frac{A^2}{12\pi G} =
 \frac{8}{3}\frac{1}{\omega+3/2}\geq 0\,,
 \quad
 \beta \equiv \frac{9}{2b_1A^2} = \frac{60\pi}{\alpha NG}\geq 0\,,
 \quad
 \nu \equiv \frac{b_2}{b_1} = - \frac {11}{18}\,.
 \]
Obviously, a particular solution exists for $\alpha = 1$ or,
equivalently, $\omega = 7/6$ which, because of ${\tilde a}'' = 0$,
i.e., ${\tilde a} = c_1 + c_2 \eta$,  is given by (a third
integration constant for later convenience is called $c_3^2$)
 \begin{eqnarray}
 \label{B1}
 a(\eta) &=& \bigg[ c_3^2(c_1+c_2\eta)^2
 +\frac{2B}{\beta c_2}(c_1+c_2\eta)
 +\frac{\nu c_2^2}{\beta}\frac{1}{(c_1+c_2\eta)^2} \bigg]^{1/2}\ ,
 \\
 \label{B2}
 \phi(\eta) &=& - \frac{3}{2A}\ln\bigg[c_3^2
 +\frac{2B}{\beta  c_2}\frac{1}{(c_1+c_2\eta)}
 +\frac{\nu c_2^2}{\beta}\frac{1}{(c_1+c_2\eta)^4} \bigg].
 \end{eqnarray}
This solution describes an expanding Universe with time-dependent dilaton
which decreases to some constant value $-(3/A) \ln c_3$;
see also Sect.~4~(a2) and Fig.~5.
Let us now consider the scalar curvature $R = 6 a''/ a^3$ of corresponding
to that solution. Using the following abbreviations
 \begin{eqnarray}
 \chi = c_3^{1/2}(c_1 + c_2 \eta),\qquad
 b = B/(\beta c_2 c_3^{3/2}), \qquad
 n = \nu c_2^2 / \beta,
 \nonumber
 \end{eqnarray}
we obtain
 \begin{eqnarray}
 R = \frac{12n}{c_3}\
 \frac{(n + 4b \chi^3 + 3 \chi^4) - b^2/2n}
      {(n + 2b \chi^3 + \chi^4)^3}\ .
 \end{eqnarray}
In the special case $B=0$, which results in $b=0$, this expression
simplifies. However, in both cases the asymptotic values for
$\eta\rightarrow\infty$ coincide,
 \begin{eqnarray}
 R_{\textrm{ as}} = - \frac{88}{15\pi} \frac{NG}{3\omega + 2}
 \frac{1}{c_2^6 c_3^5 \eta^8}\,;
 \end{eqnarray}
therefore, depending on the sign of $c_3$, the space of
negative (or positive) curvature approaches the flat one quite fast.

Second, since $v$ is not explicitly involved,
let us change the variables according to
 \[  s(\eta) = \ln a(\eta), \qquad w(\eta) = v'(\eta); \]
then, instead of Eqs.~(\ref{basic1}) and (\ref{basic2}), we obtain
 \begin{eqnarray}
  s'''  + w'' - 2(1+\nu)(s'+w)^3 +\beta w {\textrm{e}}^{2s} & = & B\ ,
  \nonumber \\
  w'  + 2s'w + w^2 + \alpha(s'' + s'{}^2) & = & 0\ , \nonumber \\
 \end{eqnarray}
or, equivalently, presupposing $\alpha\neq 0,1$,
 \begin{eqnarray}
 \label{x1}
 w'' &=&-\frac{\alpha}{1-\alpha}\Big(B - \beta w {\textrm{e}}^{2s}
         + 2\nu (w+s')^3\Big) \nonumber\\
     &&  +\frac{2w}{\alpha}\Big((1+\alpha)w^2 + w' + (2+3\alpha)w s'
       +3\alpha s'{}^2\Big)\,,\\
 \label{x2}
   s''  &=& - s'{}^2 - \frac{1}{\alpha}(w^2+w'+2ws').
\end{eqnarray}

These equations may be subjected to a symmetry group analysis
\cite{GROUP}. As a result it follows that in the general case
($B\neq 0$) only the generator of translations along $\eta$
exists:
 \begin{eqnarray} X_1 = \partial_\eta\, . \nonumber
 \end{eqnarray}
However, in the case $B=0$ an additional generator exists:
 \begin{eqnarray} X_2 = \eta\partial_\eta - w\partial_w -
 \partial_s\,. \nonumber
 \end{eqnarray}
The corresponding invariants are $\eta w \equiv W$ and $\eta \rm
e^s \equiv g$.

Choosing $W = h_1 = const., ~g = - h_2 = const.$ we get the
particular solution which is already known from \cite{GOZ}:
 \begin{equation}
 \label{goz1}
 w \equiv A \phi'(\eta)/3 = h_1/\eta,
  \end{equation}
 \begin{equation}
 \label{goz2} \textrm{e}^{s(\eta)} \equiv a(\eta) = - h_2 /\eta.
 \end{equation}
The values of $h_1$ and $h_2$ are determined by
 \begin{eqnarray}
   h_1^2  - 3 h_1 + 2 \alpha & = & 0\,, \nonumber \\
   \beta h_{2}^{2} - \left(4(1-\alpha) +
     2\nu(h_1-1)^3/h_1\right) & = & 0\,, \nonumber \\
 \end{eqnarray}
leading to
\begin{eqnarray}
 h_1 = \frac{1}{2}\left(3\pm\sqrt{9-8\alpha}\right),
 \quad
 h_{2}^{2} = \frac{1}{3\beta} \left(1
              - \frac{14}{3}\, \alpha + \frac{11}{2 \alpha }
              - \frac{11 h_1}{6\alpha} \right);
\nonumber
\end{eqnarray}
from this it is obvious that in order for $h_1$ to be real it is
necessary that $\alpha \leq 9/8$ or, equivalently, $47/54 \leq
\omega$. \footnote[5]{Note that the correct expression for $H_1=
A/3h_1$, Eq.~(22) in Ref.~\cite{GOZ}, is given by $H_1 =
-(3/16)\sqrt{4\pi G (2\omega+3)} \left(1\mp \sqrt{1-\frac{128}{27}
\frac{1}{2\omega + 3}}\right)$.} This solution, corresponding to
inflationary universe with linearly growing dilaton, has constant
scalar curvature $R = 12 / h_2^2 > 0$.

\section{Equations of motion: physical time}
\setcounter{equation}{0}

In order to be able to give a physical interpretation of the
(particular) solutions let us shift back to physical time $t$:
\begin{equation}
\label{phystime} \textrm{ d} t = a(\eta) \textrm{ d} \eta, \qquad
      \partial_\eta = a(t) \partial_t\,.
\end{equation}
Then the above Eqs.~(\ref{x1}) and (\ref{x2}) change into
 \begin{equation}
 \label{eqt}
 \begin{aligned}
  v_{tt} & + 3s_tv_t + v_t^2 + \alpha(s_{tt} + 2s_t^2) = 0\,, \\
  s_{ttt}& + v_{ttt} + 3s_t(s_{tt}+v_{tt})
         + (s_{tt}+2 s_t^2)(s_t + v_t)  \\ &
 \hphantom{=+s_{tt}}
  - 2(1+\nu)(s_t + v_t)^3 + \beta v_t  =  B {\textrm{e}}^{-3s}\,, \\
 \end{aligned}
 \end{equation}
where now
 \[
 s(t) = \ln a(t), \qquad w(t) \equiv a(t) v_t(t)\,.
 \]
This may be rewritten as follows ($\alpha \neq 1$):
 \begin{equation}
 \begin{aligned}
  0  & =   s_t + v_t - u\,, \\
  0  & =   s_{tt} + 2s_t^2 -
 \frac{1}{1-\alpha}(u_t + us_t + u^2)\,, \\
  B \textrm{e}^{-3s}  & =   u_{tt} + 3u_ts_t +
 \frac{1}{1-\alpha}u(u_t + us_t + u^2)
 -2(1+\nu)u^3 + \beta(u-s_t)\, . \\
 \end{aligned}
 \end{equation}
Obviously,  there are 5 constants required to fix any solution at
$t=t_0$; we chose them as follows:
 \begin{equation}
  \begin{aligned}
  s(t_0) & \equiv  s_0  = \ln a(t_0), \quad
     s_t(t_0)\equiv s_1 = a_t(t_0)/a(t_0), \quad
     v(t_0) \equiv v_0 = A \phi(t_0)/3,   \\
  u(t_0) & \equiv  u_0  =  v_t(t_0) + s_{t}(t_0)
             \equiv v_1 + s_1, \quad
     u_t(t_0) \equiv u_1 = v_{tt}(t_0) +  s_{tt}(t_0)
              \equiv v_2 + s_2 . \\
 \end{aligned}
 \end{equation}
\par
Let us change, by choosing $u_t = q$ and $s_t = p$, the above set
of differential equations into another one being only of first
order. Then the basic equations to be studied in the following are
given by
 \begin{eqnarray}
 \label{input}
  u_{t} &=& q \,, \qquad s_{t}=p \,,
                               \qquad v_{t}=u-p \,, \nonumber\\
  p_{t} & = & - 2p^2+\frac{1}{(1-\alpha)}(q+up+u^2)\,, \\
  q_{t} & = & -3pq -\frac{u}{(1-\alpha)} (q+up+u^2) +
        2(1+\nu)u^3+\beta(p-u)+B \textrm{e}^{-3s}\,.
\nonumber
\end{eqnarray}
The two constants, $q_0 \equiv q(t_0)$ and $p_0 \equiv p(t_0)$, which
in addition to $s_0$, $u_0$ and $v_0$ define any solution of
Eq.~(\ref{input}) are given as $u_1= q_0$ and $s_1 = p_0$.

The variable $v$ is not directly involved into these equations.
Hence, one only has to find solutions of the four equations for
the variables $s,u,p,q$, and afterwards the function $v$ is simply
obtained by quadratures. Furthermore, if $B\equiv 0$  the same is
true about $s$. Therefore, we should distinguish the cases $B=0$
and $B\neq 0$.

\subsection{Stationary solution}

The set (\ref{input}) of first order differential equations for $B=0$
has a stationary solution which is of physical relevance. It is determined
by
 \begin{eqnarray}
 \label{stationary1}
 u=u_{\textrm{st}}, \qquad p=p_{\textrm{st}}, \qquad q=0 ,
 \end{eqnarray}
where $u_{\textrm{st}}$ and $p_{\textrm{st}}$ are expressed as
follows:
 \begin{equation}
 \label{stationary2}
 p_{\textrm{st}}=\frac{u_{\textrm{st}}}{h}\,, \quad \
 u_{\textrm{st}}^2
 = \frac{\beta}{2} \frac{h(h-1)}{(1+\nu)h^2 - 1}\,, \quad \
 h=-\frac{1}{2}\left(1 \pm \sqrt{9-8\alpha} \right) \,.
\end{equation}
Substituting these values of $p_{\textrm{st}}$, $u_{\textrm{st}}$
and $q=0$ into the equations for $s$ and $v$, and integrating the
latter, we obtain the following linear dependencies
 \begin{equation}
 \label{stationary3}
 s=s_0 + (t-t_0)p_{\textrm{st}} \,, \qquad
 v=v_0 + (t-t_0)(u_{\textrm{st}}-p_{\textrm{st}})\,.
 \end{equation}
Obviously, this generalizes the solution given by
Eqs.~(\ref{goz1}) and (\ref{goz2}), which are obtained for $t_0 =
0, s_0$ and $v_0~{\rm arbitrary}, p_{\textrm{st}} = 1/h_2, u_{\rm
st} = (1-h_1)/h_2$. The solution (\ref{stationary3}) corresponds
to an exponentially increasing ($p_{\textrm{st}}>0$) universe with
a linearly growing BD dilaton and constant scalar curvature $R=12
p_{\textrm{st}}^2$.

\subsection{Numerical analysis}

In order to get some insight into the various types of behaviour
of the solutions of the system (\ref{input}) we made a systematic
numerical analysis. Here, we present only the general result.

\begin{figure}[hbtp]
\centerline{
\includegraphics[width=0.98\textwidth]{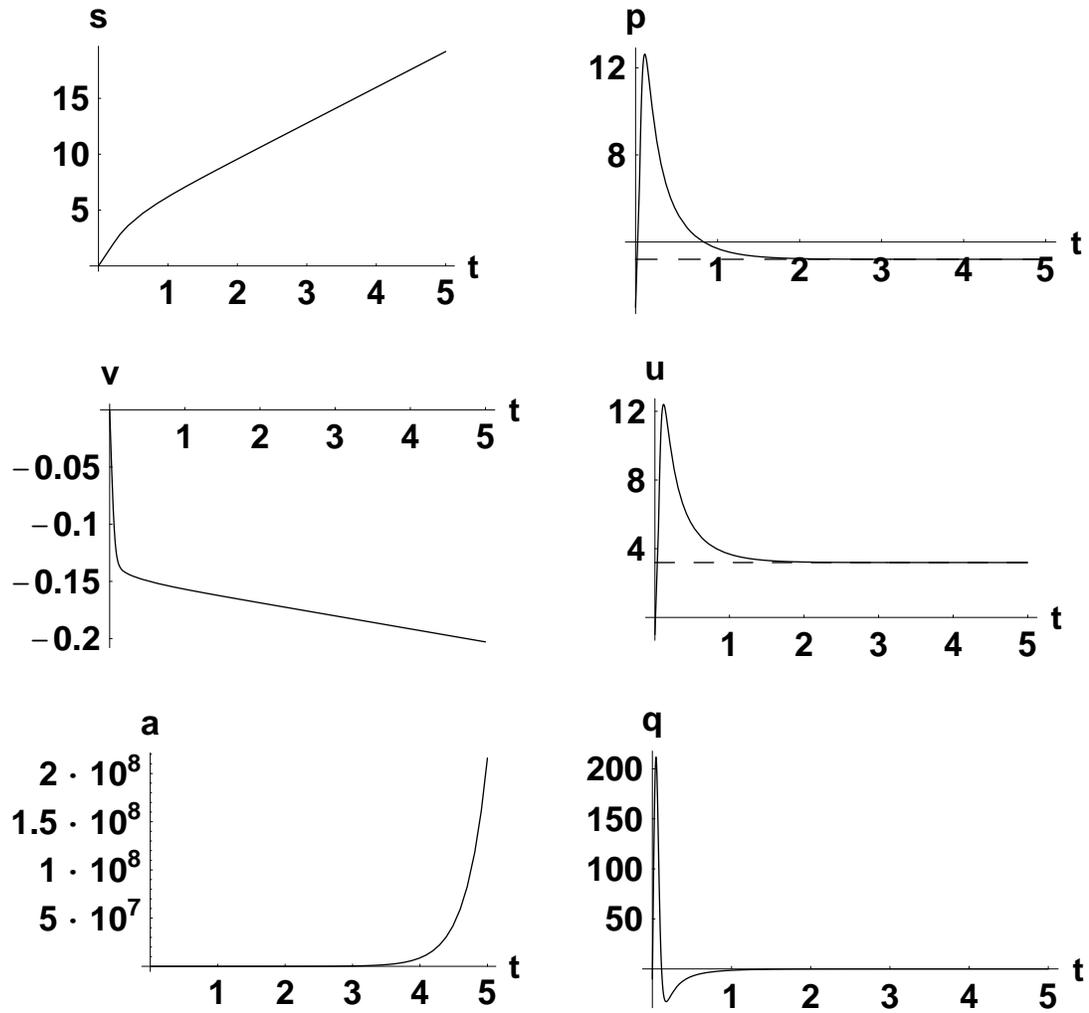}}
\caption{Dependence of functions $u$, $p$, $q$, $s$, $v$ and
$a=\exp(s)$ for $\omega = 500$, $N=10$, $B=0$ and
 $u_0=-1$, $p_0=1$, $q_0=-10$.}
 \label{Fig1}
 \end{figure}

\begin{figure}[hbtp]
\centerline{\includegraphics[width=0.98\textwidth]{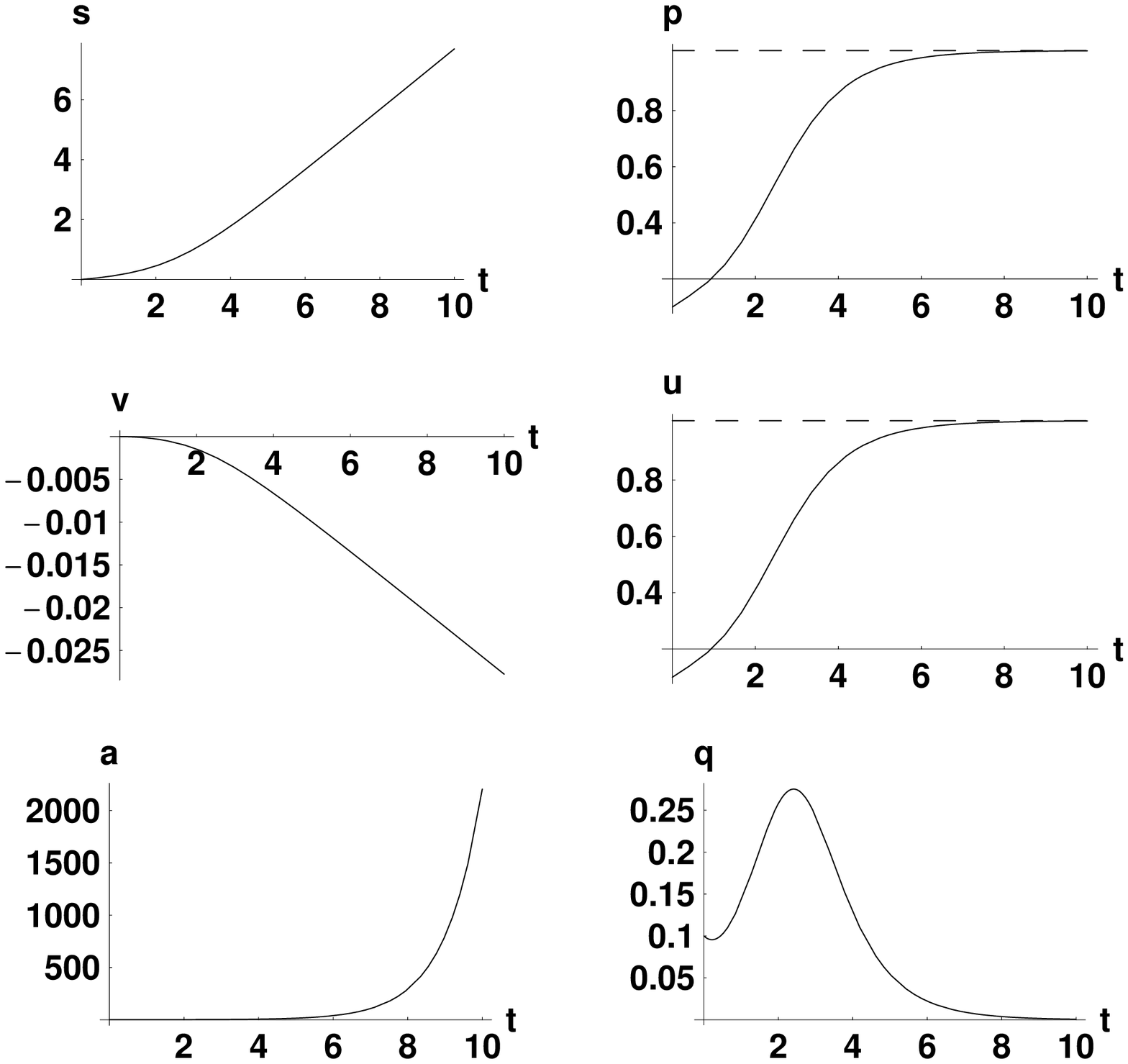}}
\caption{Dependence of functions $u$, $p$, $q$, $s$, $v$ and
$a=\exp(s)$ for $\omega = 500$, $N=100$, $B=0$ and
 $u_0=0.1$, $p_0=0.1$, $q_0=0.1$.}
 \label{Fig2}
 \end{figure}

\begin{figure}[hbtp]
\centerline{
\includegraphics[width=0.98\textwidth]{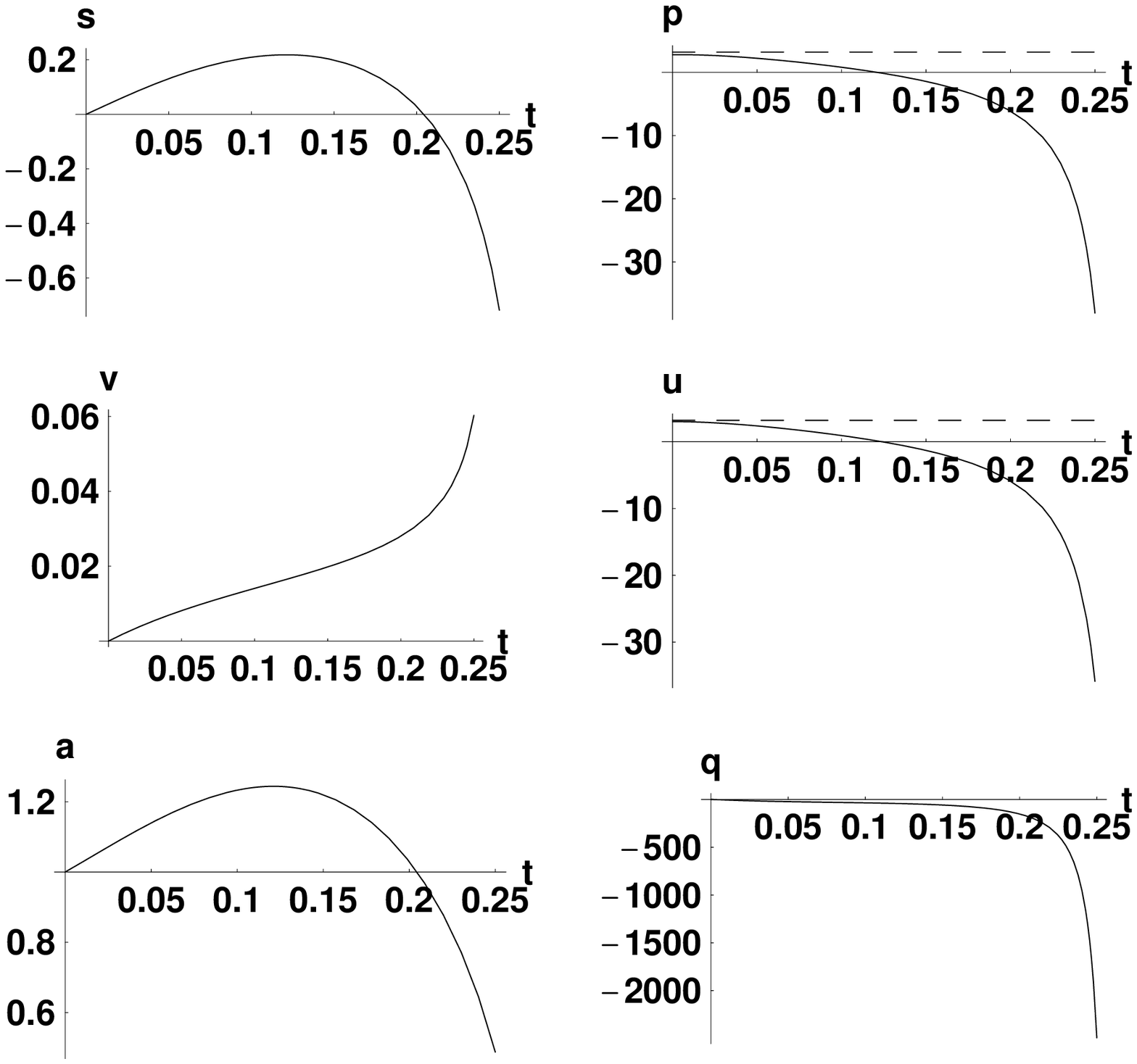}}
\caption{Dependence of functions $u$, $p$, $q$, $s$, $v$ and
$a=\exp(s)$ for $\omega = 500$, $N=10$, $B=0$ and $u_0=3$,
$p_0=2.8$, $q_0=1$.}
 \label{Fig3}
 \end{figure}

\begin{figure}[hbp!]
\centerline{
\includegraphics[width=0.99\textwidth]{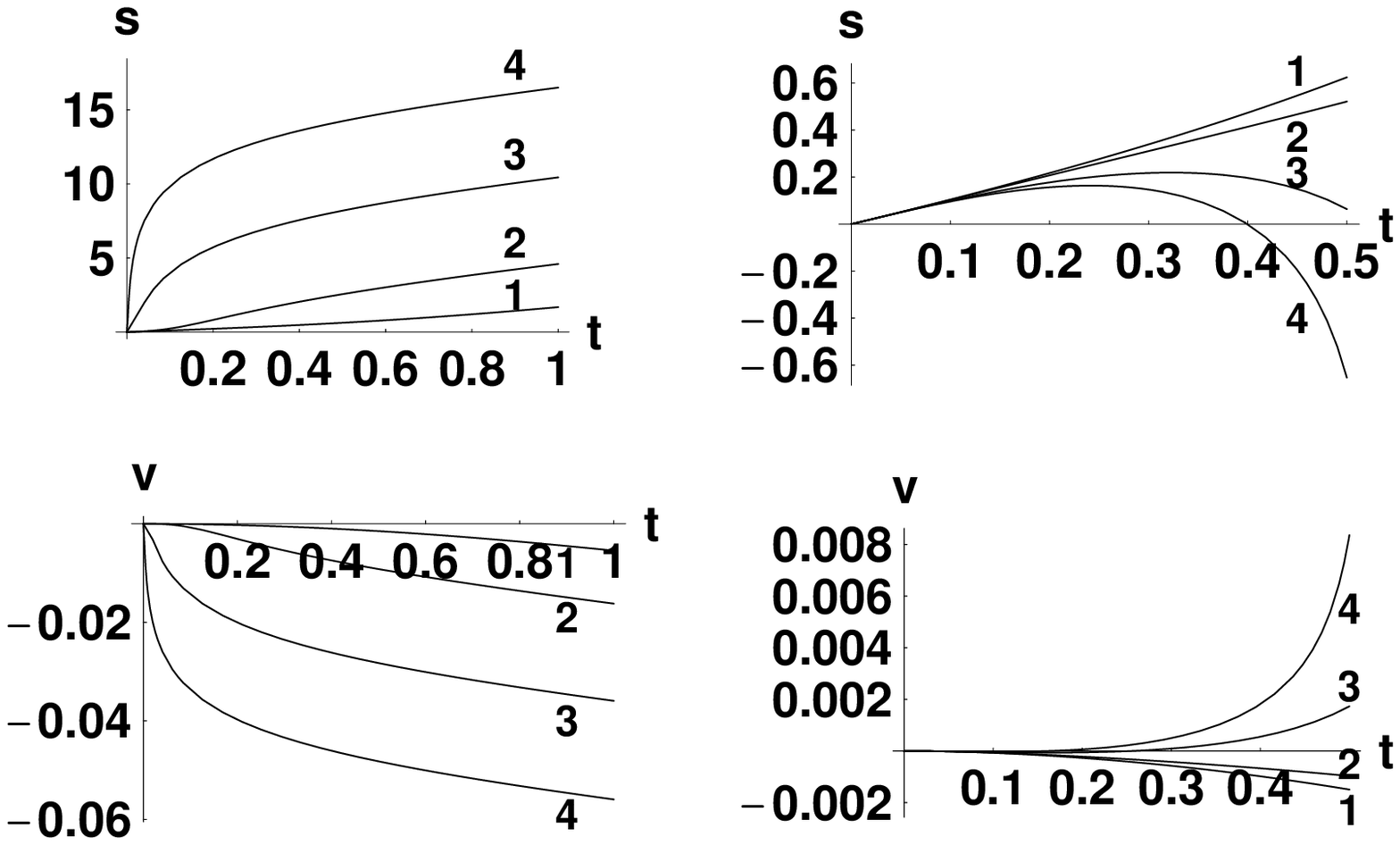}}
\caption{Dependence of functions $s$ and $v$ for $\omega = 500$,
$N=10$, $u_0=1$, $p_0=1$, $q_0=1$ and different values of $B$:
 \newline
 left panel: 1 -- $B=0$, 2 -- $B=10^3$, 3 -- $B=10^6$, 4 -- $B=10^9$.
 \newline
 right panel: 1 -- $B=0$, 2 -- $B=-10$, 3 -- $B=-40$, 4 -- $B=-60$.}
 \label{Fig4}
 \end{figure}

In the case $p_0 \geq u_0$, i.e. $v_1 \leq 0,$ and $B\geq 0$ the
solutions for $t \rightarrow \infty$ very fast approximate the
stationary solution (\ref{stationary1}) -- (\ref{stationary3}). In
these cases different values of $p_0, u_0, q_0$ and $B$ change the
shape of the solutions only quantitatively; different values of
$N$ show only a qualitative change for short times but have the
same asymptotics; cf. Figs.~\ref{Fig1} and \ref{Fig2} as well as
left panel of Fig.~\ref{Fig4}. However, if $p_0 <  u_0$, i.e.,
$v_1 > 0,$ and $B< 0$ the behaviour is quite different, cf.
Fig.~\ref{Fig3} as well as right panel of Fig.~\ref{Fig4}. In that
cases the solution shows eventually (damped) oscillations around
the asymptotic solutions or exponential increasing
(i.e.,~explosion--type) behaviour; furthermore, Fig.~\ref{Fig3}
shows a dilaton-driven collapse.

Now, we present some of the plots illustrating the typical behaviour
of solutions of the system (\ref{input}) for different values of
parameters and initial values (as in \cite{GOZ} $\omega = 500$
has been chosen). The dashed line in these plots indicate
the stationary solution.

Having so different types of numerical solutions it is
useful to discuss a possibility to construct any type of
analytical solutions. This will be presented in the next section.
Also the condition of structural stability is of great
importance; it will be considered in detail in the last section.

\section{Non-stationary solutions}
\setcounter{equation}{0}

In the input equations one can exchange the independent variable
$t$ and any of the dependent variables. For example, choosing $u$
as new independent variable we can rewrite equations (\ref{input})
in the following form
 \begin{eqnarray}
  qt_{u} & = & 1 \,, \qquad qs_{u}=p \,,\qquad
     qv_{u}   = u-p \,, \nonumber\\ \label{inputu}
  qp_{u} & = & -2p^2+\frac{1}{1-\alpha}(q+up+u^2)\, ,\\
  qq_{u} & = & -3pq -\frac{u}{1-\alpha} (q+up+u^2)+ 2(1+\nu)u^3
            + \beta(p-u)+B \textrm{e}^{-3s}\,. \nonumber
 \end{eqnarray}
\par
Instead of choosing $u$ as new independent variable we can also
choose any other dependent variable, for example $q$. This case is
of particular interest because, for $q = 0$, there exists the
stationary solution (\ref{stationary1}) of the basic equations
(\ref{input}). Therefore when $q\neq 0$ we will get the
time-dependent solution according with the first of the following
equations
 \begin{equation}
 \label{inputq}
 \begin{aligned}
   R t_{q} & =  1\,,\qquad  R s_{q} = p \,,
  \qquad R v_{q} = u-p \,, \\
   R p_{q} & =  -2p^2+\frac{1}{1-\alpha}(q+up+u^2)\,,
 \qquad R u_{q} = q \,,  \\
  {\rm where} \quad
   R  \equiv &  -3pq -\frac{u}{1-\alpha} (q+up+u^2)
 + 2(1+\nu)u^3+\beta(p-u)+B \textrm{e}^{-3s}\,. \\
 \end{aligned}
 \end{equation}
\par
In view of the polynomial dependencies of (\ref{inputu}) and
(\ref{inputq}) upon $u$ and $q$ it is possible to look for
solutions in form of an infinite series in these variables.

\subsection{Series in $u$ }

\textbf{(a)} Let us start with the particular case $B=0$. Then we
use the following representation for $q$ and $p$
\begin{equation}
\label{seru}
 q = \sum\limits_{n=0}^{\infty} A_n u^n  \,, \quad
 p = \sum\limits_{n=0}^{\infty} B_n u^n  \,,
\end{equation}
while the remaining dependencies are obtained by integrating the
first three equations in (\ref{inputu}) for $t$, $s$ and $v$.
Substitution of (\ref{seru}) into (\ref{inputu}) yields an infinite
set of equations for the coefficients $A_n$ and $B_n$:
 \begin{eqnarray}
 \label{coefu}
 &&\sum\limits_{j=0}^{k} \Big((k+1-j)A_{k+1-j}A_j
 + 3  B_{k-j}A_{j}\Big) - 2(1+\nu)\delta_{k,3}
 + \beta \delta_{k,1} - \beta B_k \nonumber\\
 && \hphantom{\sum\limits_{j=0}^{k} }
 + \frac{1}{1-\alpha}\Big(\delta_{k,3} + A_{k-1}(1-\delta_{k,0}) +
   B_{k-2}(1-\delta_{k,0})(1-\delta_{k,1})\Big) = 0 \,,
 \\
  &&
 \sum\limits_{j=0}^{k} \Big( (k+1-j) B_{k+1-j}
  A_j + 2 B_{k-j} B_j \Big)
 - \frac{1}{1-\alpha} \Big(\delta_{k,2}
 + A_k + B_{k-1} (1-\delta_{k,0})\Big) = 0 \,. \nonumber
 \end{eqnarray}
Of particular interest are those values of the parameters and
initial values when these series are truncated, i.e.,~reduce to
finite sums. In this case the requirement of vanishing for the
coefficients $A_k$ and $B_k$ for $k > k_{\textrm{ max}}$ yields
some additional conditions imposed on the coefficients $A_k$ and
$B_k$, that can be fulfilled only for some particular values of
the parameters involved. Below we present two examples of such
solutions.

\noindent
\textbf{(a1)}~~~The first example is valid for
 $\alpha=0$
 and $k_{\rm max}=2$ and is given by the formulas
 \begin{equation}
 \label{out1a}
 \begin{array}{l}
 A_0=B_0=A_1=B_2=0\,, \quad A_k=B_k=0 \,, \quad k \geq 3, \\
 B_1=1\,, \quad A_2= -1 \pm \sqrt{1+\nu}< 0\,,\\
 \mbox{} \\
 \displaystyle{
 s= s_0 - (1/A_2)\ln \left( 1-A_2 u_0 (t-t_0) \right)\,,
 \quad  v= v_0 = const. \,,}
 \\ \mbox{} \\
 \displaystyle{ q= A_2 u^2\,,  \quad
 u=p=\frac{u_0}{1-A_2 u_0 (t-t_0)}\,.}
 \end{array}
 \end{equation}
This solution confirms that different types of behavior of
solutions are possible depending upon the initial conditions. For
$u_0=p_0 >0 $ we have transition to the stationary state with zero
asymptotic values of $u$ and $p$ at $t \to \infty $, while for
$u_0<0 $ we have an explosion--type behavior whose singularity
lies at $t-t_0=1/(A_2u_0)$. The scalar curvature is positive and,
in the limit $t \rightarrow \infty$, approaches $6\, (1\pm
\sqrt{1+\nu})(1\mp\sqrt{1+\nu})^{-2} \, t^{-2} $.
\par
The solution described by formulas (\ref{out1a}) is valid for
$\alpha = 0$ and corresponds to $p=u$. However, in this case a
more general form of the behavior of functions $q$ and $t$ upon
$u$ (as compared to (\ref{out1a}) ) can be obtained that is given
as follows ($C = const. $):
 \begin{equation}
 \label{out1g}
 \begin{aligned}
{} &  t-t_0 = \mp \frac{1}{2 C } \int\limits_{r_0}^{r} \! \textrm{
d} r
 \left( \sqrt{1+\nu} - 1
 - r \right)^{-\frac{3}{4}-\frac{1}{4\sqrt{1+\nu}}}
 \left(\sqrt{1+\nu} + 1 + r \right)^{-\frac{3}{4}+\frac{1}{4\sqrt{1
 +\nu}}} \,,  \\
 {} &  s = s_0 + \frac{1}{4 \sqrt{1+\nu}} \ln
    \left( \frac{ \sqrt{1+\nu} + 1 + r }
   { \sqrt{1+\nu} - 1 - r } \right)  \,,  \quad
   v= v_0 = const. \,, \\
 {} & q= u^2 r\,, \quad  u= p = \pm C
   \left( \sqrt{1+\nu} - 1 - r \right)^{-\frac{1}{4}
  + \frac{1}{4\sqrt{1+\nu}}} \left( \sqrt{1+\nu} + 1
  + r \right)^{-\frac{1}{4}-\frac{1}{4\sqrt{1+\nu}}} \,. \\
 \end{aligned}
 \end{equation}
The result of the previous case (\ref{out1a}) arises if we impose
the restriction $r_u=0$ that leads to constant values of $r=-1 \pm
\sqrt{1+\nu}$.
\par
Obviously, as long as $p=u$ these solutions are quite
special since they correspond to a constant dilaton.
Furthermore, because of $p=u$ the solution does not depend
on $\beta$ (and $G$)!
\par
\noindent \textbf{(a2)}~~~The second example is valid for $\alpha
= 1$, $k_{\textrm{ max}}=4$, and is given by the
formulas\footnote[6]{For $\alpha = 1$ it seems more convenient to
substitute the representation (\ref{seru}) directly into
Eq.~(\ref{eqt}).}
\begin{equation}
\label{out1b}
\begin{aligned}
{} &  A_0=B_0=A_1=B_2=A_3=B_4=0\,, \quad A_k=B_k=0  \,,
     \quad k \geq 5,  \\ &
   B_1=1\,, \quad A_2= -2 \,, \quad A_4=-B_3=2\nu / \beta \,,
     \\ {} &
  t-t_0=\frac{1}{2u} - \frac{1}{2u_0} +
    \frac{1}{2\sqrt{-\beta / \nu}}\left( \arctan
    \frac{u}{\sqrt{-\beta /\nu }} -\arctan\frac{u_0}
    {\sqrt{-\beta/\nu}} \right) \,,  \\ {} &
  s= s_0 - \frac{1}{4}\ln \left( \frac{ u^2 (u^2-\beta /\nu)}
             {u_0^2 (u_0^2-\beta / \nu)} \right)\,, \quad
  v= v_0 + \frac{1}{2}\ln
  \left( \frac{u^2-\beta /\nu}{u_0^2-\beta /\nu} \right)  \,,  \\ {} &
  p = u \left( 1-\frac{2 \nu }{\beta}\, u^2 \right) \,, \quad
  q = -2u^2 \left( 1-\frac{\nu }{\beta}\, u^2 \right) \,. \\
 \end{aligned}
 \end{equation}
\par
This solution may be obtained also directly from Eqs.~(\ref{B1})
and (\ref{B2}) where we used the conformal time $\eta$. Indeed,
putting $B=0$ in (\ref{B1}) and substituting into (\ref{conftime})
we get
 \[ \textrm{d} t = \frac{1}{4}\,\frac{ \textrm{d} q}{q}
    \sqrt{({\nu}/{\beta})+ ( c_{3}^2/c_{2}^{2} ) q },
    \quad q=\left( c_{1}+c_{2} \eta \right)^4 \,,
 \]
which is easily integrated as
 \[
    t = const. + \frac{1}{2} \left(
    \sqrt{{\nu}/{\beta} + ({c_{3}^2}/{c_{2}^{2}})q }
    - \sqrt{- {\nu}/{\beta}}
     \arctan  \sqrt{-1 - ({c_{3}^2}/{c_{2}^{2}})(\beta/\nu)q }
    \right)\,.
 \]
Making use of the definition of $u=v_t+s_t=a_{\eta}/a^2 +
v_{\eta}/a$ we get, after differentiation of (\ref{B1}) and
(\ref{B2}), the following relation between $u$ and $a$:
 \[ u^{-1}= {(c_1+c_2 \eta) a(\eta)}/c_2 =
    \sqrt{({\nu}/{\beta}) + ({c_{3}^2}/{c_{2}^{2}})q }\,.
 \]
 Eliminating $q$ with the help of the last formula from the
 expression for $t$ we get
 \[ t= const. +\frac{1}{2} \left(
     \frac{1}{u} - \sqrt{-{\nu}/{\beta}}
     \arctan \frac{1}{\sqrt{-\nu /\beta}u} \right) \,,
 \]
which coincides with (\ref{out1b}) if we take into account
$\arctan x = \pi /2 - \arctan(1/x)$. The rest of
Eqs.~(\ref{out1b}) may be checked immediately.

In the limit $(t - const.) \to 0 $ the last formula gives $u \sim
(t-const.)^{-1/3}\, $ ; in the opposite limit $ t/ \sqrt{-\nu
 /\beta} >> 1 $ we have $u \sim (1/2)\left(t - const. +
(\pi/4 )\sqrt{-\nu/\beta}\right)^{-1}$.
\par
Even in the case of $B \neq 0$, when we can hardly find the
analytical dependence of $t$ upon $\eta$, the use of the formula
(\ref{conftime}) can help to find the dependence of $t$ upon
$\eta$ by quadratures:
 \[
 t = const + \int\frac{\textrm{d} p}{p} \sqrt{({\nu}/{\beta})+
     ( 2 B /\beta c_{2}^{3} ) p^3 +
     ( c_{3}^2/c_{2}^{2} ) p^4 }, \quad
     p=\left( c_{1}+c_{2} \eta \right) \,,
 \]
and  formulas (\ref{B1}), (\ref{B2}) have the form
 \[ a = (c_2/p) \sqrt{({\nu}/{\beta})+
           ( 2 B /\beta c_{2}^{3} ) p^3 +
           ( c_{3}^2/c_{2}^{2} ) p^4 }, \quad v = \ln (p/a)\,.
 \]
\par
\begin{figure}[htbp]
\begin{center}
\includegraphics[width=0.9 \textwidth]
{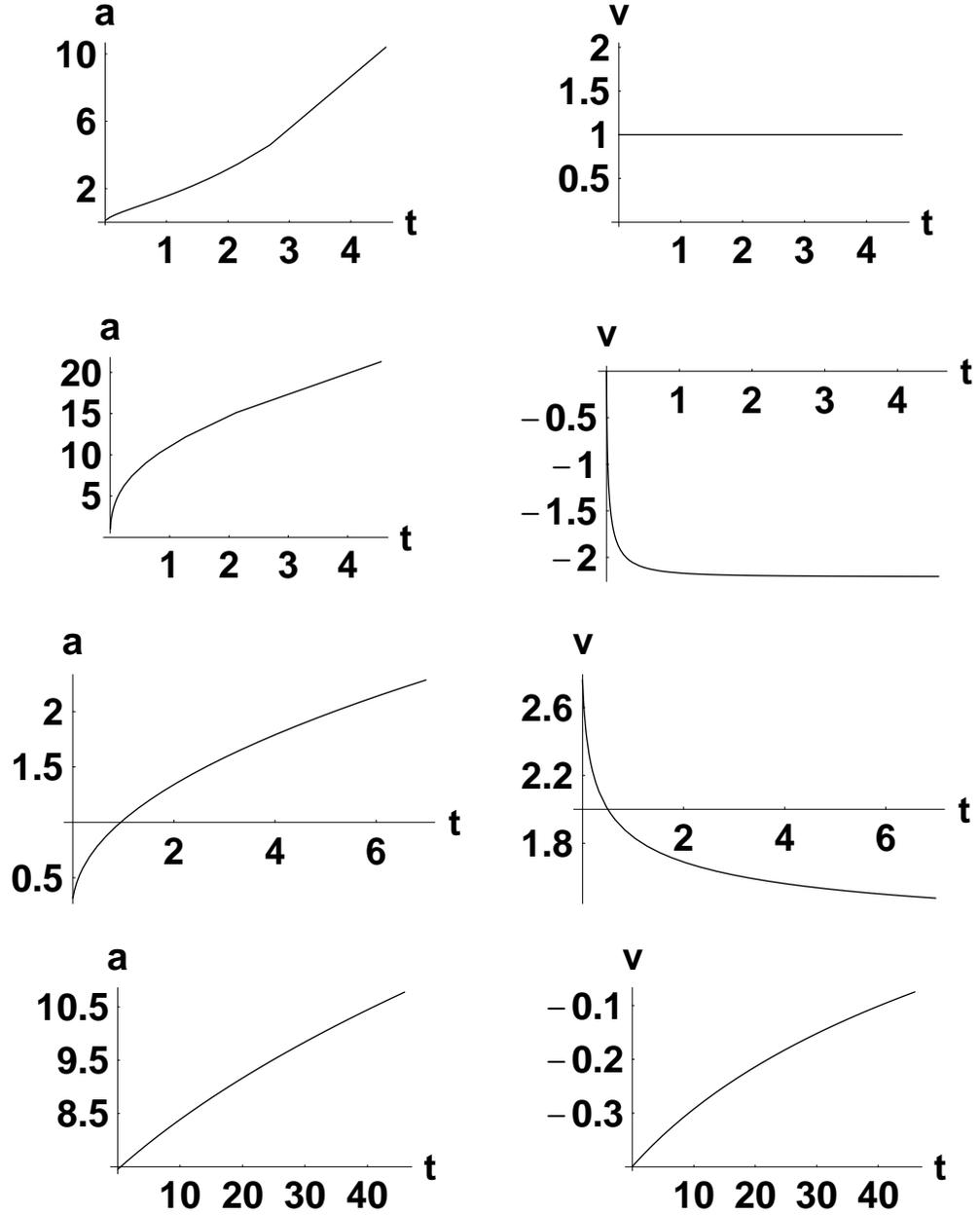}
\end{center}
\caption{Characteristic behavior of $a(t)$ and $v(t)$ resulting
 for $B=0$ from
 (a) equations \protect{(\ref{out1g})} for $C=1$ and $v_0=1$ (first, top
 panel),
 (b) equations \protect{(\ref{out1b})} for $N=100, G=1$ (second panel);
  and from equations \protect{(\ref{B1},\ref{B2})} for $N=100,
  c_2=1,\, c_3^2=0.1$ with (c) $B=10$ (third panel) and (d) $B=-0.45$,
 (bottom, fourth panel), respectively.}
 \label{Fig5}
\end{figure}

The behavior of $a$ and $v$ upon $t$, resulting from
Eqs.~(\ref{out1g}) for $B=0$ as well as (\ref{B1}), (\ref{B2}) for
$B\neq 0$ will be obtained by numerical integration. Some
characteristic plots, related to these equations and
(\ref{out1b}), are shown in Fig.~\ref{Fig5}.

\noindent
\textbf{(b)} In the general case when $ B\neq
0$ we use the following representation for $q$, $p$ and $s$
\begin{equation}
\label{seru1}
 q = \sum\limits_{n=0}^{\infty} A_n u^n  \,, \quad
 p =  \sum\limits_{n=0}^{\infty} B_n u^n  \,, \quad
 s = s_0-\frac{1}{3}\ln\Big(\sum\limits_{n=0}^{\infty} C_n u^n \Big)\,,
\end{equation}
while the remaining dependencies are obtained by integrating the
equations in (\ref{inputu}) for $t$ and $v$. Substitution of
(\ref{seru1}) into (\ref{inputu}) yields an infinite set of
equations for the coefficients:
 \begin{equation}
 \label{coef2}
 \begin{aligned}
 {} &
 \sum\limits_{j=0}^{k}\Bigl( (k+1-j) A_{k+1-j}A_j
  + 3 B_{k-j} A_j \Bigr)
 \\ {} &
  \hphantom{\sum\limits_{j=0}^{k} (k+1-j) }
 - 2(1+\nu)\delta_{k,3} + \beta \delta_{k,1} - \beta B_k - B e^{-3s_{0}} C_k
   \\ {} &  \hphantom{\sum\limits_{j=0}^{k} (k+1-j)}
   + \frac{1}{1-\alpha}\left( \delta_{k,3}
   +  A_{k-1}(1-\delta_{k,0})
   +  B_{k-2}(1-\delta_{k,0})(1-\delta_{k,1}) \right) = 0 \,,
     \\ {} &
   \sum\limits_{j=0}^{k} \Bigl( (k+1-j)B_{k+1-j} A_j
   + 2 B_{k-j}B_j \Bigr)
     \\ {} &  \hphantom{\sum\limits_{j=0}^{k} (k+1-j)}
   - \frac{1}{1-\alpha} \left( \delta_{k,2}
   + A_k + B_{k-1} (1-\delta_{k,0}) \right)  = 0 \,,
     \\ {} &
     \sum\limits_{j=0}^{k} \Bigl((k+1-j) C_{k+1-j}A_j
    +3 C_{k-j}B_j \Bigr) =0 \,. \\
  \end{aligned}
 \end{equation}
Again, we present two examples of solutions that correspond to
truncated series (\ref{seru1}) for some $k>k_{\rm max}$.

\noindent
 \textbf{(b1)}~~~The first solution corresponds to
$\alpha =0$, $k_{\textrm{ max}}=2$
 \begin{equation}
 \label{out2a}
 \begin{aligned} {} &
   A_0=B_0=C_0=A_1=C_1=B_2=C_2=0 \,, \quad
      A_k=B_k=C_{k+1}=0 \,, \quad k \geq 3 \,,  \\ {} &
    B_1=1\,, \quad A_2= -1 \,, \quad C_3=1 \,,  \\ {} &
  s = - (1/3)\ln(-7 u_0^3/9B) + \ln \left( 1+ u_0 (t-t_0)
        \right) \,,  \quad   v= v_0 = const. \,,  \\ {} &
   u = p = \frac{u_0}{1 + u_0 (t-t_0)} \,, \quad
  q= - \left( \frac{u_0}{1 + u_0 (t-t_0)}\right)^2 \,.
 \end{aligned}
 \end{equation}

Here again we see the effect of explosive behavior for $u_0<0$ at
$t-t_0 \to -(1/u_0)$ and asymptotic stability at $t \to \infty $
in the opposite limit $u_0 > 0$ similar to the observation above.
Again, the curvature is positive and approaches $ 6 / t^2$.
\par
\noindent
 \textbf{(b2)}~~~The second solution corresponds to some $\alpha \neq 0$,
$k_{\rm max}=2$:
\begin{equation}
\label{out2b}
\begin{aligned} {} &
  A_0=B_0=C_0=A_1=B_2=C_2=0\,, \quad
     A_k=B_k=C_{k}=0 \,, \quad k \geq 3,  \\ {} &
  B_1=\pm (1/2) \sqrt{1+\nu} \,, \quad A_2= -3B_1=
     \mp (3/2)\sqrt{1+\nu}  \,,  \\ {} &
   C_1=\frac{\beta}{B}(1-B_1) \exp \left( 3s_0 \right) \,,
     \quad  \alpha = \left( \frac{1-B_1}{B_1}\right)^2 \equiv
           1+4\frac{1 \mp \sqrt{1+\nu}}{1+\nu} \,,  \\ {} &
   s= - (1/3)\ln(\beta(1-B_1)u_0/B) +  \frac{1}{3}
         \ln \left( 1+ 3B_1 u_0 (t-t_0) \right)\,,\\ {} &
  v= v_0  +  \frac{1-B_1}{3B_1} \ln \left( 1+ 3B_1 u_0
   (t-t_0) \right)\,, \quad u= \frac{u_0}{1 + 3 B_1 u_0 (t-t_0)}\,,
    \\ {} &
   q= -3B_1 \left( \frac{u_0}{1 + 3B_1 u_0 (t-t_0)}\right)^2\,,
    \quad  p= B_1 \frac{u_0}{1 + 3 B_1 u_0 (t-t_0)}\,. \\
 \end{aligned}
 \end{equation}
Here, the system demonstrates explosive behavior for $B_1 u_0 <
0$, while for $B_1 u_0 >0$ we have vanishing $u$, $q$ and $p$ for
$t \to \infty $, and the raise of $s$ and $v$ is softer than in
the stationary case. Here, the curvature is negative and
approaches $ - (2/3)\, t^{-2}$.

\subsection{Series in $q$ }
Here we will examine only the particular case of $B=0$. Then we
use the following representation for $u$ and $p$
 \begin{equation}
 \label{serq}
 u = \sum\limits_{n=0}^{\infty} A_n q^n  \,, \quad
 p = \sum\limits_{n=0}^{\infty} B_n q^n  \,, \quad
 \end{equation}
while the remaining dependencies are obtained by integrating the
first three equations in (\ref{inputq}) for $t$, $s$ and $v$.
Substitution of the expansions (\ref{serq}) into the next two
equations in~(\ref{inputq}) yields the following infinite set of
equations for the coefficients:
\begin{equation}
 \label{coefq}
 \begin{aligned} {} &
 \sum\limits_{j=0}^{k} (k+1-j)A_{k+1-j}R_j
 - \delta_{k,1}  = 0 \,,  \\ {} &
  \sum\limits_{j=0}^{k} \Bigl( (k+1-j) B_{k+1-j}R_j
  + 2 B_{k-j} B_j \Bigr) \\ {} &
  \hphantom{  \sum\limits_{j=0}^{k} \Bigl( (k+1-j) B_{k+1-j}R_j}
  - \frac{1}{1-\alpha} \Big(\delta_{k,1}
  + \sum\limits_{j=0}^{k} A_j \left(  A_{k-j} + B_{k-j} \right)
  \Bigr) = 0 \,,  \\
 \end{aligned}
 \end{equation}
\begin{equation}
 \begin{aligned}
  {\textrm{ where}} \quad {R_j} & =
  (\delta_{j,0} - 1)
   \left(3 B_{j-1}+ \frac{1}{1-\alpha} A_{j-1}\right)
   + \beta ( B_{j} - A_{j} )   \\
 {}  &  +  \sum\limits_{i=0}^{j} \sum\limits_{l=0}^{i} A_{l}A_{i-l}
   \left( 2 (1+\nu) A_{j-1}
   -  \frac{1}{1-\alpha}\left( B_{j-1} + A_{j-1} \right) \right)
   \,.\\
  \end{aligned}
 \end{equation}
Despite being of principal interest this set of equations is quite
complicated. There exists one evident truncation of the series
(\ref{serq}) namely when only zero order terms are taken into
account, $A_0 \neq 0$, $B_0 \neq 0$, whilst $A_k = B_k = 0$ for $k
\geq 1$. This truncation corresponds to $q=0$ and the stationary
solution (\ref{stationary3}). Unfortunately, in the case $q\neq 0$
we were not able to find a nontrivial truncation of these series.

\section{Stability analysis}
\subsection{Small perturbations}
Here, we consider only the case $B=0$. Then the basic equations
have stable solutions of the form (\ref{stationary1}). By
linearizing these equations with respect to small perturbations
$\delta u = u-u_{\textrm{ st}}$, $\delta q = q$, $\delta p =
p-p_{\textrm{ st}}$ we obtain a system of linear equations that
admit solutions of the form $\delta u \propto U\, \textrm{
e}^{\lambda t}$, $\delta q \propto Q\, \textrm{ e}^{\lambda t}$
and $\delta p \propto P \, \textrm{ e}^{\lambda t}$ where
$\lambda$ obeys the characteristic equation
 \begin{eqnarray}
 \label{character}
  \left[ \lambda^2 + \lambda \left( 3p_{\textrm{ st}}
    + \frac{u_{\textrm{ st}}}{1-\alpha}\right)
    + \frac{u_{\textrm{ st}}}{1-\alpha}(2p_{\textrm{ st}}+3u_{\textrm{ st}})
    + \beta -6 (1+ \nu) u_{\textrm{ st}}^2 \right] \nonumber \\
   \times \left( \lambda + 4p_{\textrm{ st}}-\frac{u_{\textrm{ st}}}
      {1-\alpha}\right) - \frac{1}{1-\alpha}\left(\beta
     -\frac{u_{\textrm{ st}}^2}{1-\alpha}\right)
     (\lambda + p_{\textrm{ st}}+2u_{\textrm{ st}}) = 0\,.
 \end{eqnarray}
Below, on Fig.~{\ref{Fig6}}, we present a plot illustrating the
dependence of real parts of solutions of the equation
(\ref{character}) upon the parameter $\omega$. The graphic shows
that there exist stable solutions in a large region of the
parameter $\omega$.
\begin{figure}[hbt]
\begin{center}
\includegraphics[width=0.98\textwidth]{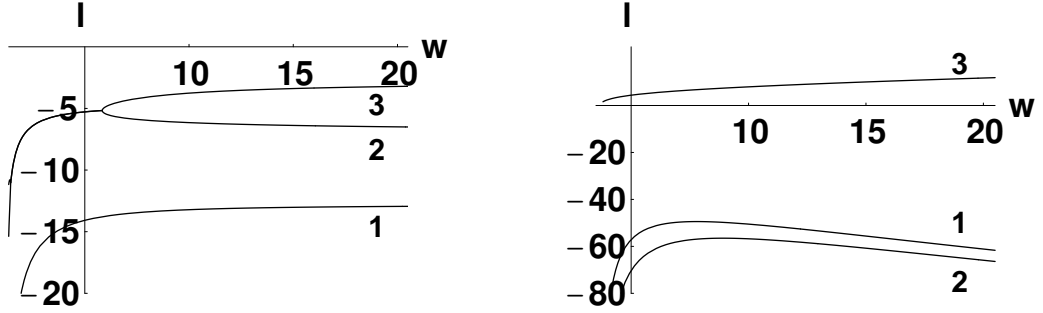}
\end{center}
\caption{Dependence of real parts of solutions of the
characteristic equation \protect{(\ref{character})} upon $\omega$
that demonstrates the existence of stable (curves $1, \ 2$ and $3$
on the left panel correspond to negative real parts) and unstable
(the curve $3$ on the right panel corresponds to the positive real
part) stationary solutions.} \label{Fig6}
\end{figure}

\begin{figure}[hbt]
\begin{center}
\includegraphics[width=0.98\textwidth]{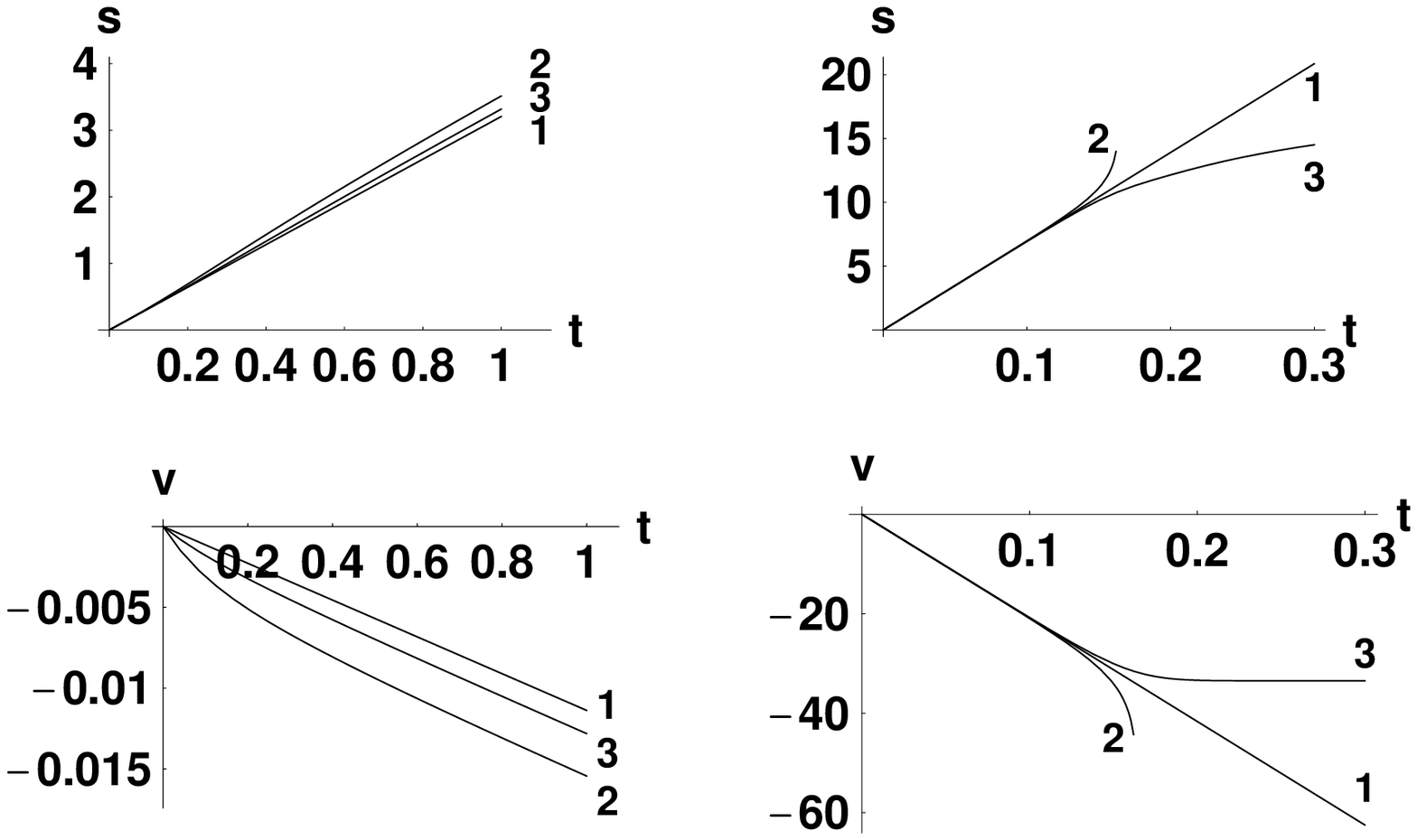}
\end{center}
 \caption{Dependence of functions $s$ and $v$ for $\omega = 500$, $N=10$
 for stable (left panel) and unstable (right panel) solutions.}
 \label{Fig7}
\end{figure}

The corresponding numerical solutions of input equations
(\ref{input}) on Fig.~{\ref{Fig7}} shows  the behavior of the
functions $s$ and $v$ in the vicinity of the stationary state.
Straight lines (curves 1) correspond to stationary solutions
(\ref{stationary1}) with $u_{\textrm{ st}}=3,19549$, $p_{\textrm{
st}}=3,20687$ for the left panel and $u_{\textrm{st}}=-138,843$,
$p_{\textrm{ st}}=69,5447$ for the right panel. For the left panel
curves 2 relate to $u_0-u_{\textrm{ st}}=-0.02$, $p_0-p_{\textrm{
st}}=0.01$ and $q_0=0.01$; curves 3 relate to $u_0-u_{\rm
st}=-0.02$, $p_0-p_{\textrm{ st}}=-0.01$ and $q_0=-0.01$. For the
right panel curves 2 relate to $u_0-u_{\textrm{ st}}=-0.02$,
$p_0-p_{\textrm{st}}=-0.01$ and $q_0=0.01$; curves 3 relate to
$u_0-u_{\textrm{st}}=0.01$, $p_0-p_{\textrm{ st}}=-0.01$ and
$q_0=-0.01$.

\subsection{Transition to the stationary solution}

It is evident that for initial values that are close enough to the
stable stationary values we can get an analytical solution for
arbitrary initial values $u(t=0)=u_0$, $p(t=0)=p_0$ and
$q(t=0)=q_0$. This solution is given by linear combinations of
exponential functions defined by solutions of the characteristic
equations presented in the previous subsection with coefficients,
depending upon the initial values
 \begin{equation}
 \label{transition1}
  \begin{aligned} {} &
  q = D_1 \exp({\lambda_1 t})
       + D_2 \exp({\lambda_2 t})
       + D_3 \exp({\lambda_3 t})\,,  \\ {}&
  u = u_{\textrm{st}} + (D_1/\lambda_1) \exp({\lambda_1 t})
            + (D_2/\lambda_2) \exp({\lambda_2 t})
            + (D_3/\lambda_3) \exp({\lambda_3 t})\,,  \\ {}&
   p = p_{\textrm{st}}  + D_1\Lambda_1 \exp({\lambda_1 t})
             + D_2\Lambda_2 \exp ({\lambda_2 t})
             + D_3\Lambda_3 \exp({\lambda_3 t})\,,  \\ {}&
   {\rm with} \qquad
  \Lambda_i= \frac{(p_{\textrm{st}}+2u_{\textrm{st}}+\lambda_i )/\lambda_i}
   {(4p_{\textrm{st}}+\lambda_i)(1-\alpha)-u_{\textrm{st}}} \,. \\
 \end{aligned}
 \end{equation}
Here the constants $D_1$, $D_2$ and $D_3$ are expressed in terms
of the initial values as follows:
 \begin{equation}
 \label{transition2}
 \begin{aligned} {} &
  D_1 = \frac{ \Delta_1}{\Delta_0} \,, \quad
  D_2 = \frac{\Delta_2 }{\Delta_0 }\,, \quad
  D_3 = \frac{\Delta_3}{\Delta_0 } \,,  \\
  \end{aligned}
 \end{equation}
 \[
 \begin{aligned} {} &
 \Delta_0= \Lambda_1\left( \frac{1}{\lambda_3}-\frac{1}
  {\lambda_2}\right) +\Lambda_2\left( \frac{1}{\lambda_1}
  -\frac{1}{\lambda_3}\right)+\Lambda_3\left( \frac{1}
  {\lambda_2}-\frac{1}{\lambda_1}\right) \,,  \\ {}&
 \Delta_1= (u_0-u_{\textrm{st}})\left( {\Lambda_2}-{\Lambda_3}\right)
  + q_0\left(\frac{\Lambda_3}{\lambda_2}-\frac{\Lambda_2}
  {\lambda_3}\right) + (p_0-p_{\textrm{st}})\left( \frac{1}
  {\lambda_3}-\frac{1}{\lambda_2}\right) \,,  \\ {}&
 \Delta_2= (u_0-u_{\textrm{st}})\left( {\Lambda_3}-{\Lambda_1}\right)
  +q_0\left(\frac{\Lambda_1}{\lambda_3}-\frac{\Lambda_3}
  {\lambda_1}\right) +(p_0-p_{\textrm{st}})\left( \frac{1}
  {\lambda_1}-\frac{1}{\lambda_3}\right) \,,  \\ {}&
 \Delta_3= (u_0-u_{\textrm{st}})\left( {\Lambda_1}-{\Lambda_2}\right)
    +q_0\left(\frac{\Lambda_2}{\lambda_1}-\frac{\Lambda_1}
    {\lambda_2}\right) +(p_0-p_{\textrm{st}})\left( \frac{1}
    {\lambda_2}-\frac{1}{\lambda_1}\right)
  \,. \\
 \end{aligned}
 \]
Integrating the remaining equations we get the following formulas
for $s$ and $v$ that describe the transition to the stationary
state
 \begin{equation}
 \label{transition3}
 \begin{aligned}
  s  =  s_0 & + (t-t_0) p_{\textrm{st}}
    + D_1 \ \frac{\Lambda_1}{\lambda_1} \left( \textrm{e}^{\lambda_1 t} -
    \textrm{e}^{\lambda_1 t_0} \right)   \\ {} &
   + D_2 \ \frac{\Lambda_2} {\lambda_2 }\left( \textrm{e}^{\lambda_2 t} -
    \textrm{e}^{\lambda_2 t_0} \right)
   + D_3 \ \frac{\Lambda_3}{\lambda_3 } \left( \textrm{e}^{\lambda_3 t}-
    \textrm{e}^{\lambda_3 t_0} \right) \,,  \\
  v  =  v_0 & + (t-t_0) ( u_{\textrm{st}}-p_{\textrm{st}})
         + \frac{D_1}{\lambda_1 }  \left(\frac{1}
         {\lambda_1} -{\Lambda_1} \right)
      \left( \textrm{e}^{\lambda_1 t}- \textrm{e}^{\lambda_1 t_0} \right)
   \\ {} &
   + \frac{D_2}{\lambda_2 } \left(\frac{1}{\lambda_2}
    -{\Lambda_2} \right)  \left( \textrm{e}^{ \lambda_2 t}
    - \textrm{e}^{\lambda_2 t_0} \right)
  + \frac{D_3}{\lambda_3 } \left( \frac{1}{\lambda_3}
  -{\Lambda_3} \right) \left( \textrm{e}^{\lambda_3 t}
  - \textrm{e}^{\lambda_3 t_0} \right)\,. \\
  \end{aligned}
  \end{equation}
It is clearly seen that the asymptotic behaviour of the functions
$s$ and $v$ depends linear upon $t$, and the slope of these lines
is defined by the stationary values, while the slope at the point
$t=0$ is defined by the initial values.

\begin{figure}[hbtp]
\begin{center}
\includegraphics[width=0.98\textwidth]{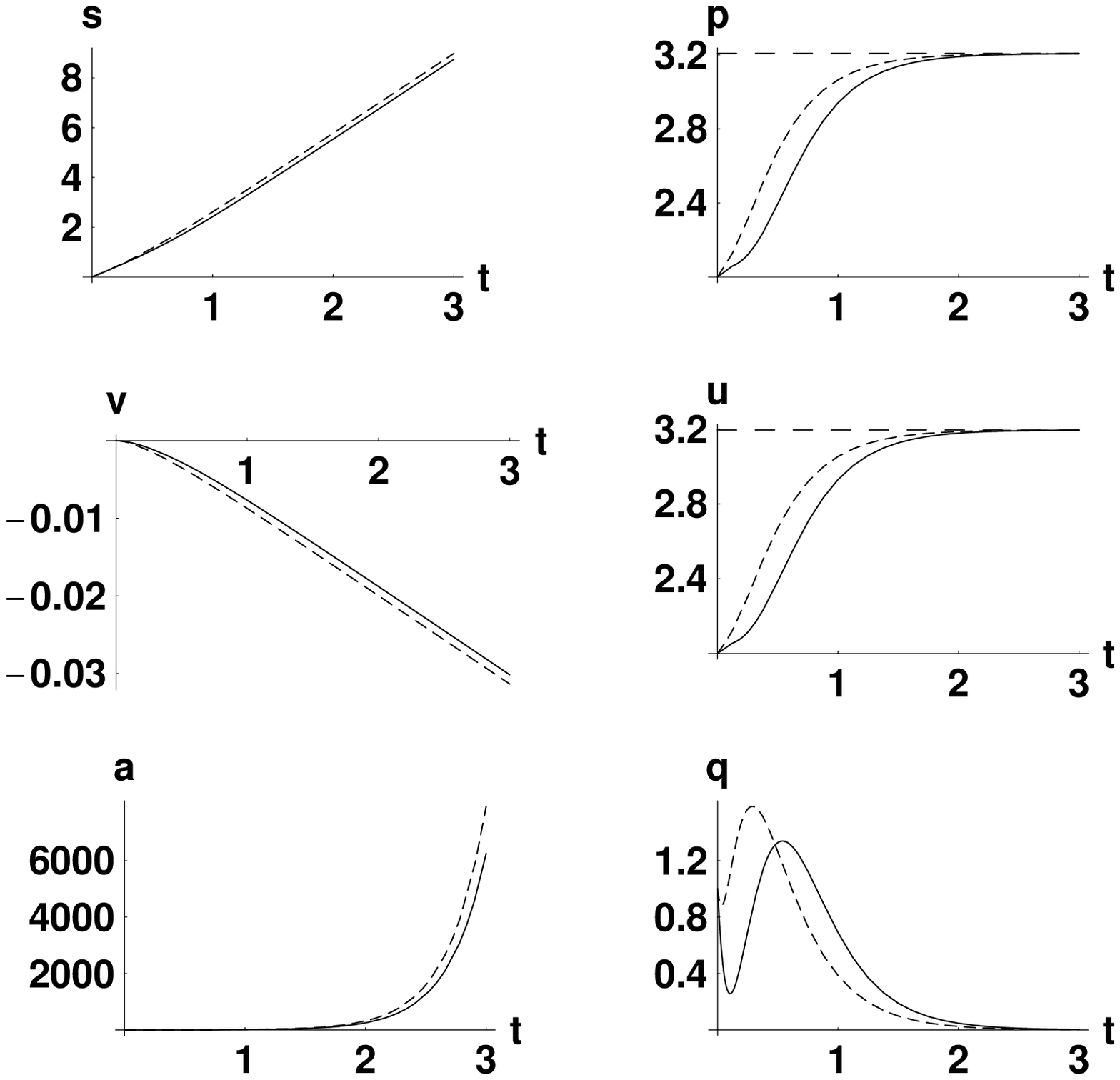}
\end{center}
\caption{Exact numerical (solid line) and approximate analytical
\protect{(\ref{transition3})} (dashed lines) solutions of basic
equations \protect{(\ref{input})} for $\omega = 500$, $N=10$,
$B=0$, $q_0=1$ and initial values $u_0=p_0=2$ close to stationary
ones $u_{\textrm{st}}=3.19549$, $p_{\textrm{st}}=3.20687$.}
\label{Fig8}
\end{figure}

In order to compare the perturbative solutions (\ref{transition3})
with the exact ones  of the basic equations (\ref{input}) we
considered the behaviour of $s$, $v$, $u$, $p$, $q$ and $a$ for
initial values $u_0$, $p_0$ and $q_0$ that are close to the
stationary values $u_{\textrm{st}}$, $p_{\textrm{st}}$ and $q=0$
for $\omega = 500, B = 0$ and different values of $N$. An example,
for $N=10$, is presented in Fig.~\ref{Fig8}.

These plots demonstrate that there is only a small discrepancy between
exact and approximate solutions for physically interesting quantities
$a$ (resp. $s$) and $v$, whereas the difference for $u,p$ and $q$ is
somewhat larger. These differences decrease for increasing values of $N$.
Obviously, the exact solution is approximated very fast.

\begin{figure}[hbt]
\begin{center}
\includegraphics[width=0.98\textwidth]{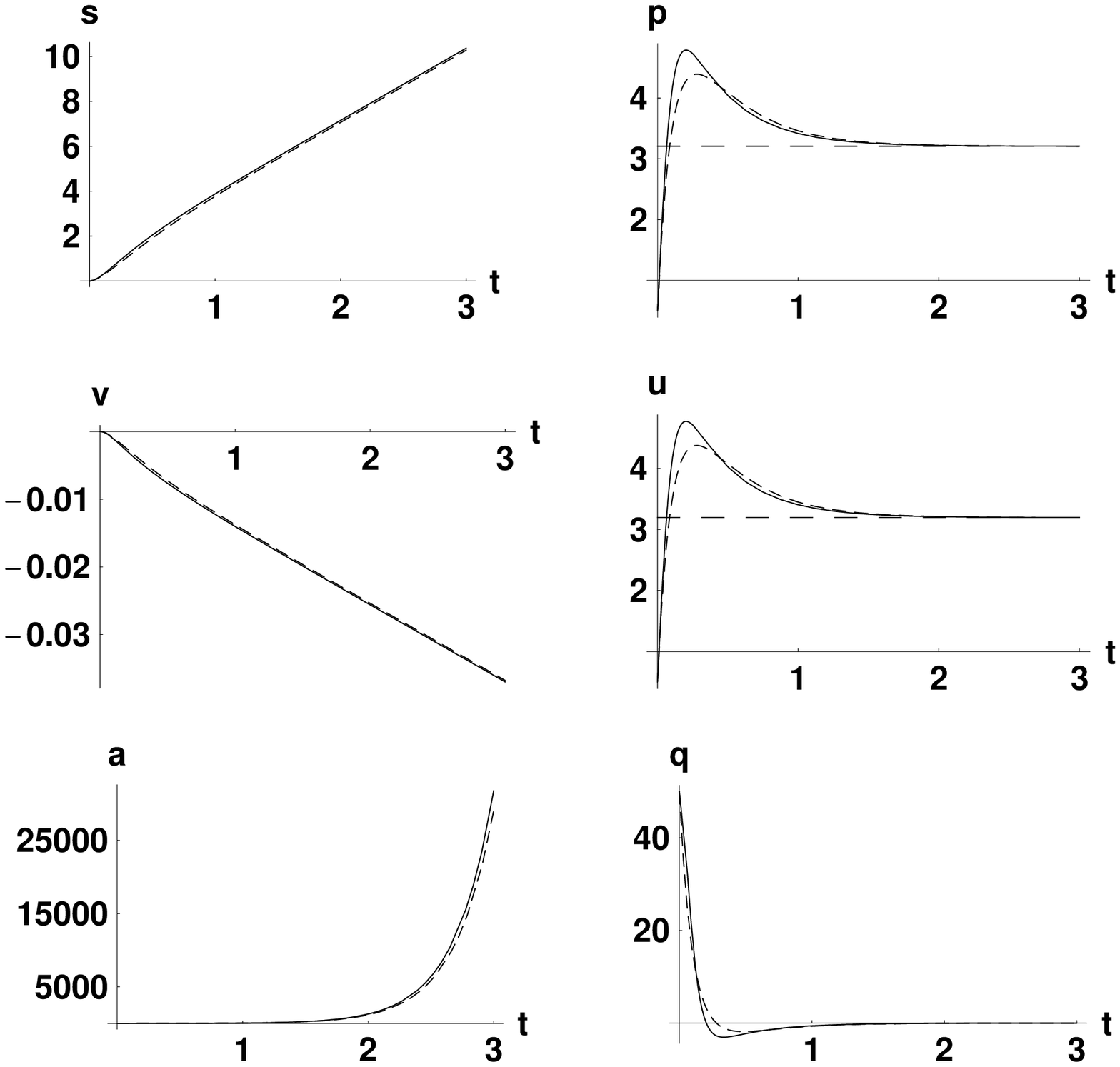}
\end{center}
\caption{Exact numerical (solid line) and approximate analytical
\protect{(\ref{transition3})} (dashed lines) solutions of basic
equations \protect{(\ref{input})} for $\omega = 500$, $N=10$,
$B=0$, $q_0=50$ and initial values $u_0=p_0=0.5$ not close to
stationary ones $u_{\textrm{st}}=3.19549$,
$p_{\textrm{st}}=3.20687$.} \label{Fig9}
\end{figure}

But, even if the initial values differ substantially from the
stationary ones the approximate formulas still give a satisfactory
accuracy for $u_0=p_0$. Moreover, as the input equations are
linear in $q$ it appears that the approximate solution is
sometimes valid even when the initial value of $q_0$ is not small.
This is clearly seen in Fig.~\ref{Fig9}.

\afterpage{\clearpage}

Also in these cases the stationary solution is approximated after
a few units of time.

\section{Conclusion}
Analysing the evolution equations in conformal time we were able
to present a new exact solution (for $\alpha = 1 \sim \omega =
7/6$). After transforming to physical time we found `stationary'
solutions of the corresponding basic set of first order
differential equations (restricted to $B=0$) which generalize the
inflationary solution already known from an earlier study
\cite{GOZ}. The numerical analysis of the basic set showed quite
different types of dilaton-driven evolution in cases $B=0$ as well
as $B\neq0$: In a wide range of the parameters of the theory the
stationary solution is approximated very fast; furthermore, there
are also quite different non-stationary solutions showing
explosion type, oscillatory as well as collapsing behaviour.
Seeking solutions as (finite) power series we were able to find
additional particular solutions of the basic set with a (modified)
logarithmic raise of $s(t)$ which, in comparison with the
stationary solution raising linearly, lead to a more soft
inflation; however, nontrivial dilatonic behaviour results only in
the cases (a2) and (b2). Finally, we presented a stability
analysis for the case $B=0$ thereby showing how the transition to
the stationary solution occurs.

It is very interesting that the mathematical methods applied here
to study the conformally flat Universe with time-dependent dilaton
solutions leading to effective forth order equations of motion are general
enough to be used also in various related contexts.
As an immediate application we may consider the problem of quantum
creation and annihilation of Anti de Sitter Universe and Anti de Sitter
black holes due to quantum effects of (dilaton coupled) matter.
Such a phenomenon has been investigated recently in Ref. \cite{BO}.
In these works the anomaly induced effective action for dilaton coupled
spinors has been used similarly to the present paper. Moreover,
the Anti de Sitter Universe in conformally flat coordinates looks
very similar to the de Sitter (inflationary) Universe: the corresponding
equations of motion are - with some change of signs - almost
the same as Eqs.~(\ref{GOZ1}) and (\ref{GOZ2}) (where, however, the role
of time is played by the radial coordinate in AdS Universe).
Hence, our results may be used in the construction of other variants
of the asymptotically AdS solutions which have been found in Ref.~\cite{BO}.

\bigskip

\noindent
{\bf\large Acknowledgement}\\
The authors are very much indebted to S.D. Odintsov for numerous
stimulating discussions and a careful reading of the manuscript.
V.F. Kovalev acknowledges for a grant of Saxonian Ministry of Sciences and
Arts.


\section*{References}
\begin{thebibliography}{99}

\bibitem{GOZ}
B. Geyer, S.D. Odintsov and S. Zerbini, {\sl Phys. Lett B}
\textbf{460} (1999) 58.

\bibitem{FGT} V. Faraoni, E. Gunzig and P. Nardone,
{\it Conformal transformations in classical gravitational theories
and in cosmology}, \textit{Fundamentals of Cosmic Physics}
\textbf{20} (1999) 121.

\bibitem{will} C.M. Will, \textit{Theory and Experiments in Gravitational
Physics}, Cambridge, 1993.

\bibitem{polch} J. Polchinski, \textit{String Theory},
 Cambridge, 1998.

\bibitem{la} D. La and P.J. Steinhardt, {\sl Phys. Rev. Lett.}
 \textbf{62} (1989) 376; \\
E.W. Kolb, D. Salopek and M.S. Turner,
 {\sl Phys. Rev} \textbf{D42} (1990) 3925; \\
for a review, see: \\
 E. Kolb and M. Turner, \textit{The Very Early Universe},
New York, 1994.

\bibitem{barrow} J.D. Barrow, {\sl Phys. Rev.} \textbf{D59} (1999)  ;
      \\ J.D. Barrow and J. Magueijo,
    \textit{Class. Quant. Grav.} \textbf{16} (1999) 1435.

\bibitem{DEA} S. Nojiri and S.D. Odintsov,
 \textit{Phys. Rev.} \textbf{D57} (1998) 2363;
 \textit{Phys. Lett.} \textbf{B426} (1998) 29;
 \textbf{B444} (1998) 92;\\
S. Ichinose and S.D. Odintsov,
 \textit{Nucl. Phys.} \textbf{B539} (1999) 634;\\
P. van Nieuwenhuizen,  S. Nojiri and S.D. Odintsov, \textit{Phys.
 Rev.} \textbf{D60} 084014.

\bibitem{GROUP}
Nail H. Ibragimov, {\it Elementary group analysis and ordinary
differential equations}, John Wiley \& Sons, Chichester-Weinheim,
1999.

\bibitem{BO}
I. Brevik and S.D. Odintsov, hep-th/9912032, Phys. Lett. (to appear)\\
S. Nojiri, S.D. Odintsov, and S. Zerbini, hep-th/0001192.

\end{thebibliography}
\end{document}